# U-Net-based Surrogate Model for Evaluation of Microfluidic Channels

**Le Quang Tuyen[1], Pao-Hsiung Chiu[1], and *†Chinchun Ooi[1]**

[1]Department of Fluid Dynamics, Institute of High Performance Computing, Singapore.

*Presenting author: ooicc@ihpc.a-star.edu.sg
†Corresponding author: ooicc@ihpc.a-star.edu.sg

**Abstract**

Microfluidics have shown great promise in multiple applications, especially in biomedical diagnostics and separations. While the flow properties of these microfluidic devices can be solved by numerical methods such as computational fluid dynamics (CFD), the process of mesh generation and setting up a numerical solver requires some domain familiarity, while more intuitive commercial programs such as Fluent and StarCCM can be expensive. Hence, in this work, we demonstrated the use of a U-Net convolutional neural network as a surrogate model for predicting the velocity and pressure fields that would result for a particular set of microfluidic filter designs. The surrogate model is fast, easy to set-up and can be used to predict and assess the flow velocity and pressure fields across the domain for new designs of interest via the input of a geometry-encoding matrix. In addition, we demonstrate that the same methodology can also be used to train a network to predict pressure based on velocity data, and propose that this can be an alternative to numerical algorithms for calculating pressure based on velocity measurements from particle-image velocimetry measurements. Critically, in both applications, we demonstrate prediction test errors of less than 1%, suggesting that this is indeed a viable method.

## Introduction

Microfluidics have been applied to a variety of areas, especially in the area of biomedical diagnostics and cell separations [1-5]. Certain common features within these devices are the placement of solid columns, such as in the case of deterministic lateral displacement devices for cell separation, or for flow field sculpting [6, 7]. While it is possible to do a parametric optimization of these designs via actual experiments, these experiments can be difficult and time-consuming. Numerical methods such as computational fluid dynamics (CFD) can thus be useful as a means of narrowing the design space prior to any actual fabrication, and are commonly used as part of the design process [8, 9].

In addition, it can often be difficult to obtain full knowledge of pressure fields within microfluidic devices, even as pressure is typically used as a means of active control of valves within such devices [10]. In particular, while experimental methods such as particle image velocimetry (PIV) can be applied to obtain velocity field, other numerical methods have to be developed for calculating the pressure fields from the measured velocities [11-13].

Hence, in this work, we further explore the use of data-driven machine learning methods for the above-mentioned two potential use cases in microfluidics: 1) predicting velocity and pressure fields for new microfluidic channel designs; and 2) predicting pressure fields from velocity field information. In particular, developments in methods such as neural networks have shown great promise as a surrogate model across multiple engineering domains in recent literature [14-16]. The convolutional neural network (CNN) architecture in particular, has been applied to several fluid dynamic problems with some success, with the U-Net model being a particularly promising choice [17, 18]. Importantly, we evaluate the sensitivity of the U-Net architecture to the choice of normalization parameter for the target outputs in this work, and show that the U-Net architecture can indeed be applied to the prediction of flow and pressure fields with root mean square errors below 1%. In addition, we also propose that the use of a dual model system, whereby we create a secondary model to predict the normalization parameter, can further improve the accuracy of the model with regards to the full pressure field.

**Methods**

*Microfluidic Flow Scenarios*

For this work, the microfluidic scenario as depicted in Figure 1 was studied, representing a continuously repeating array of solid pillars in a microfluidic channel. The individual pillars extend for 10 μm, while the design parameters are assumed to be the gap between the pillars (G), and the lateral dimension of the individual small pillars (S). In total, simulations were run for S ranging from 10 to 30 μm in steps of 2 μm and G ranging from 20 to 50 μm in steps of 2 μm to create a total dataset of 176 individual flow scenarios.

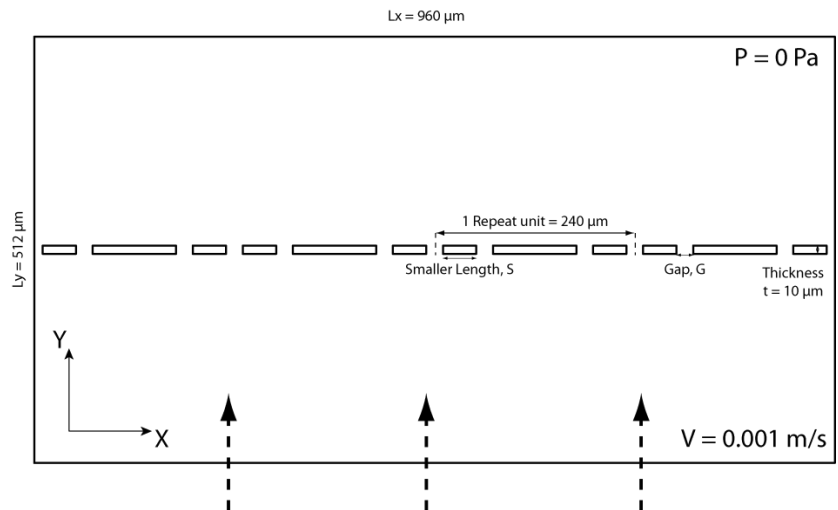

**Figure 1. Schematic of microfluidic channel scenarios studied. Two parameterizations are studied in this work: 1) width of small pillar (S), and 2) width of inter-pillar gap (G).**

The Dirichlet input boundary condition is stipulated as a uniform velocity ($V_{in}$) of magnitude 0.001 m/s, while a Dirichlet boundary condition of P = Pa is applied at the outlet. Periodic boundary conditions are specified on the two sides to simulate a continually repeating array of pillars in the flow. The steady-state velocity and pressure fields were obtained based on a pressure-projection scheme, with second order spatial discretization. Sample plots of the velocity fields obtained for 1) S = 10 μm and G = 20 μm, and 2) S = 30 μm and G = 50 μm are provided in Figure 2.

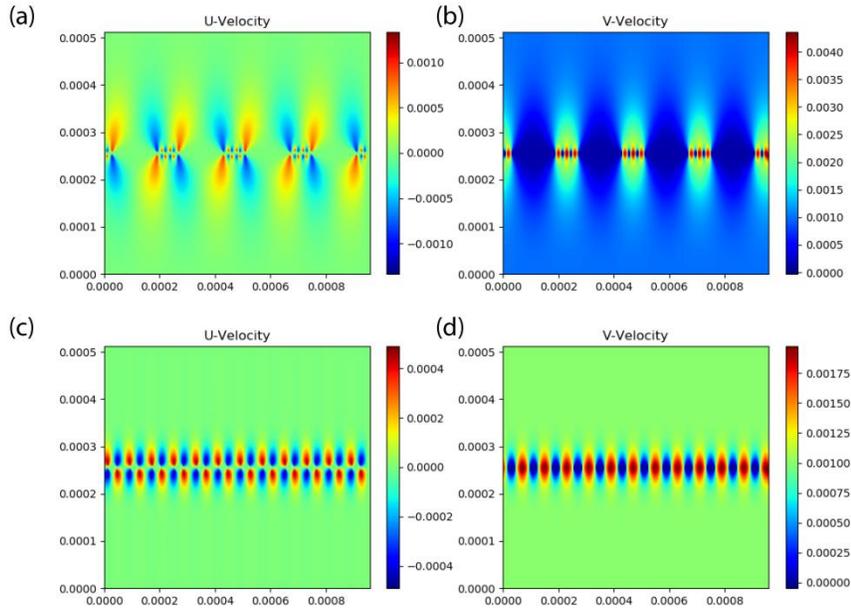

**Figure 2.** (a) and (b) are the U and V velocity fields obtained by CFD for the flow scenario with S = 10 μm and G = 20 μm while (c) and (d) are the U and V velocity fields obtained by CFD for the flow scenario where S = 30 μm and G = 50 μm.

*U-Net Architecture and Model Training*

Based on prior positive results in literature, the U-Net architecture was selected for use in this work, as illustrated in Figure 3 [17]. The model was created in Python, and all network mathematical operations used were as implemented in the base Keras and Tensorflow packages. Briefly, the U-Net has a bowtie structure, comprising separate encoder and decoder halves. Each half is made up of sequential layers that comprise of a convolutional layer, a batch normalization layer and a non-linear activation component, with the choice of activation function being the leaky ReLU (0.2) in the encoder half and the ReLU function in the decoder half. Default network hyper-parameters were used based on prior work by Thuruey et al., with no further optimization, although a convolutional kernel of size 8x8 was used instead [17]. The Adam optimizer was used for training over 10k epochs, with a learning rate of 0.0004, and 'mean squared error' (MSE) as the loss function. In each of the following models, the dataset was split into a training dataset consisting of 150 cases, while a random subset of 26 cases was used as the test set for evaluation.

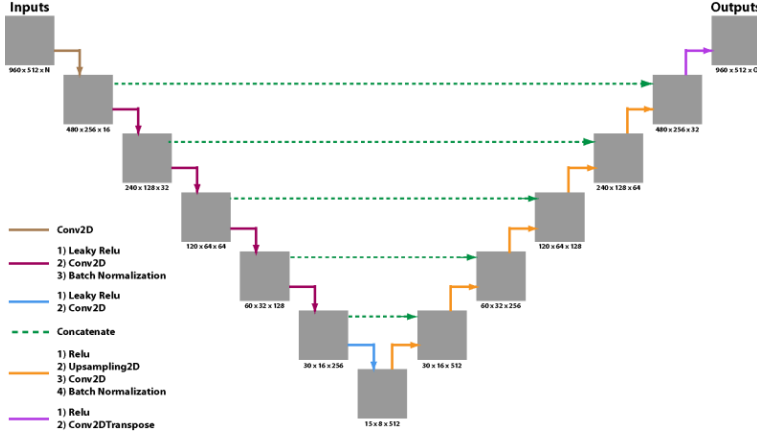
**Figure 3. Schematic of the U-Net CNN architecture as implemented in this work. Different color arrows in the schematic represent the operations that were used at each stage.**

In order to enhance predictive performance, pre-processing of the data is crucial. Prior literature in the machine learning domain typically also recommends normalization of the input and output as the gradient descent-based optimization process can be hindered by extremes in input or output value magnitude [19, 20]. Hence, a few different normalization methods for the input and output data are evaluated in this work.

*U-Net Model Prediction for Pressure based on Velocity*

For the first scenario, we assume that the velocities are provided, and the pressure field is to be predicted (Models A1, A2 and A3). Hence, the x and y velocity fields (U and V) are specified as inputs, while the output field is pressure (P). Utilizing typical scales that would be used for non-dimensionalization of the Navier-Stokes equations, we elected to normalize the input velocity fields by the magnitude of the input velocity ($V_{in}$).

Inputs for Models A1, A2 and A3:
$$U_{norm}^i = \frac{U^i}{V_{in}} \quad (1)$$
$$V_{norm}^i = \frac{V^i}{V_{in}} \quad (2)$$

Similarly, for Model A1, we used a scaling factor commonly used for non-dimensionalizing pressure in creeping flows, and further de-meaned the output pressure field by subtracting a 'typical' pressure profile as obtained by averaging across all the training data.

Output for Model A1:
$$P_{norm}^i[:,y] = \frac{L}{\mu V_{in}} * (P^i[:,y] - \frac{1}{n_{train}}\sum_{j=1}^{n_{train}} P^j[:,y]) \quad (3)$$

where the superscripts i and j represent the individual flow scenarios, and y represents the yth coordinate of the domain. L and μ represent the length scale and the fluid viscosity respectively, and were defined as L = 0.001 m, and $\mu = 0.001 \, Pa \cdot s$.

In addition, a second choice for renormalizing the pressure was by re-scaling the individual output pressures for every parametric flow scenario by the expected pressure difference between the input and output planes of the flow domain. Hence, we created a regression function for the average pressure difference due to the choices of S and G, and used that as a normalization function instead.

Output for Model A2:

$$P^i_{norm}[:,:] = \frac{P^i[:,:]}{f(S,G)} \quad (4.1)$$

$$f(S,G) = e^{5.9205} S^{0.093691} G^{1.931052} \quad (4.2)$$

where i represents the individual flow scenario, and f(S,G) is the regression function obtained for the dependence of pressure on S and G, as obtained by least-squares regression.

Lastly, as a further proof-of-concept, in most typical microfluidic set-ups, the input and output pressures are typically easy to measure, or are actual control parameters for operation. Hence, in such experiments, we anticipate being able to use actual measurement values for the pressure differential between the inlet and outlet planes. Thus, we assumed the average pressure along the inlet plane of the fluid domain is known from experiments, and used that as a normalization factor in Model A3.

Output for Model A3:

$$P^i_{norm}[:,:] = \frac{P^i[:,:]}{\frac{1}{w}\sum_{k=1}^{w} P^i[k,1]} \quad (5)$$

where i represents the individual flow scenario, and w represents the fluid cells along the inlet plane of the fluid domain.

*U-Net Model Prediction for Velocity and Pressure Fields*

For the second scenario, we assume that both velocity and pressure fields are unknown, although the geometry of the domain is specified. Hence, the following input and output fields are used instead:

Inputs for Models B1 and B2:

$$U_{input} = \begin{cases} 0, & solid\ domain \\ 0, & fluid\ domain \end{cases} \quad (6)$$

$$V_{input} = \begin{cases} 0, & solid\ domain \\ 1, & fluid\ domain \end{cases} \quad (7)$$

$$VOF_{input} = \begin{cases} 1, & solid\ domain \\ 0, & fluid\ domain \end{cases} \quad (8)$$

The input velocities are normalized to be of values 0 or 1, rather than the characteristic input velocity, for similar reasons to the normalization of input velocities in Models A1 to A3. In addition, we add an input parameter, VOF, which is the inverse of the Volume-of-Fluid parameter that is commonly used in immersed boundary methods to demarcate solid and fluid domains. In particular, the VOF parameter is assumed to be 1 when the component overlays a solid object in the domain, and is 0 otherwise. While the encodings for velocity and VOF appear similar, and the input field for U appears redundant in that it is uniformly 0 in this instance, the two input fields have been retained for generalizability across future scenarios, whereby the velocity boundary conditions might be non-zero in both spatial axes.

A similar normalization was chosen for the output values for both velocity and pressure, whereby we re-scaled the velocity and pressure fields to optimally shift their range to be between 0 and 1 or -1 and 1. Two separate models were compared, with a slight difference observed for the two instances.

Outputs for Model B1:

$$P^i_{norm}[:,:] = \frac{P^i[:,:]}{\frac{1}{w}\sum_{k=1}^{w} P^i[k,1]} \tag{9}$$

$$U^i_{norm} = \frac{U^i}{V_{in}} \tag{10}$$

$$V^i_{norm} = \frac{V^i - V_{in}}{3 * V_{in}} \tag{11}$$

Outputs for Model B2:

$$P^i_{norm}[:,:] = \frac{P^i[:,:]}{\frac{1}{w}\sum_{k=1}^{w} P^i[k,1]} - 0.5 \tag{12}$$

$$U^i_{norm} = \frac{U^i}{V_{in}} \tag{13}$$

$$V^i_{norm} = \frac{V^i - V_{in}}{2 * V_{in}} \tag{14}$$

In particular, we vary the normalization parameters across Models B1 and B2 in order to further evaluate the impact of the normalization on the predictive accuracy of the model.

**Results**

*U-Net Model Prediction for Pressure based on Velocity*

After training, the root-mean-square errors (RMSEs) of the 3 models (A1, A2 and A3) are calculated on the test set, and the respective errors are presented in Figure 4. Sample pressure predictions are also presented in Figure 5. Based on the results from models A1 and A2, we noticed that the RMSE was significantly improved in variance by using the pressure difference between the input and output planes of the microfluidic component as a scaling parameter as compared to the average of the data-set. More importantly, when the pressure difference is accurately defined, as in Model A3, the U-Net model test RMSE is also best minimized with an average value of about 0.7%.

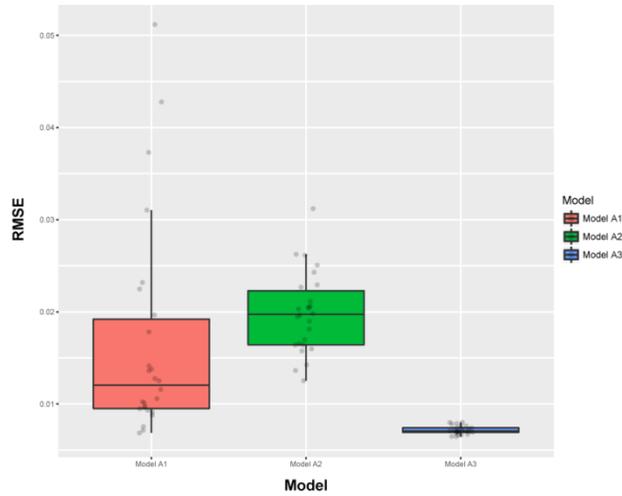

**Figure 4. Test RMSEs for the 3 different ways of normalizing the pressure output field as defined by models A1 to A3.**

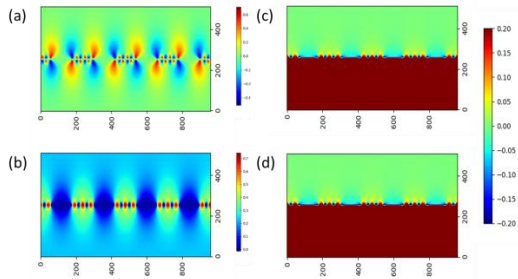

**Figure 5. Sample U, V and P contour plots as obtained for this work for a flow scenario with G = 30 μm and S = 16 μm. (a), (b) and (c) are the U, V and P contour plots as obtained from CFD, while (d) is the pressure contour plot as obtained by the U-Net Model A3.**

*U-Net Model Prediction for Velocity and Pressure Fields*

Similarly, two different normalization models are tested, and their test RMSEs are plotted in Figure 6, while sample velocity and pressure contour plots as obtained by CFD and Model B2 are plotted in Figure 7. The results indicate that the choice of normalization can also impact the accuracy of the model, and indicate that Model B2 performs slightly better, with average RMSEs of 0.5% for pressure and 0.7% for velocities.

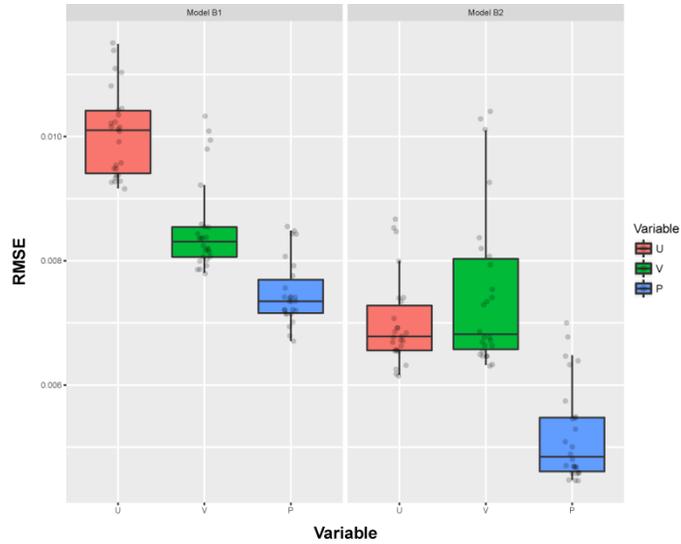

**Figure 6. Test RMSEs for the 2 different ways of normalizing the velocity and pressure output fields as defined by models B1 to B2.**

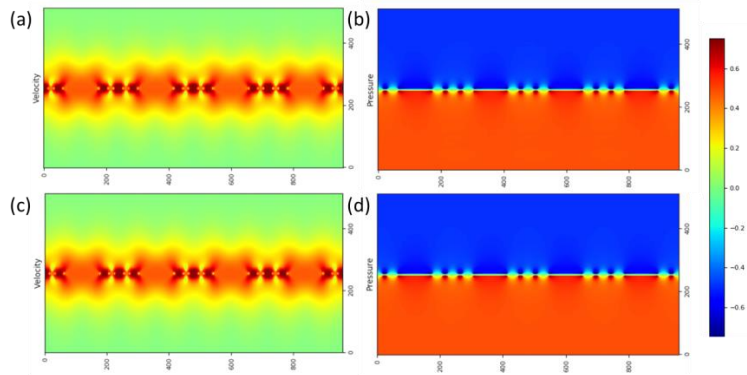

**Figure 7. Sample velocity magnitude and pressure contour plots as obtained by CFD and via our U-Net (CNN) Model B2 as presented in this work for the flow scenario with G = 34 μm and S = 14 μm. (a) and (b) are obtained by CFD while (c) and (d) are obtained by Model B2.**

## Conclusions

In this work, we demonstrate the application of a U-Net-based model to rapidly predict the pressure fields for a set of microfluidic channel designs, when velocity fields are known, such as in the case of PIV experiments, and also, for the prediction of both velocity and pressure fields when new designs are to be evaluated. We demonstrate that we can obtain prediction errors of less than 1% with an appropriate choice of normalization parameters, and also highlight that the normalization parameters can improve model prediction.

While the microfluidic designs here are only a subset of many more complicated possibilities, this proof-of-concept work shows the potential for utilizing deep learning methods, and U-Net specifically, as a surrogate model for design evaluation and optimization in microfluidic applications. We anticipate that this method would be of interest to microfluidic experimentalists, as an easy to use surrogate model for rapid evaluation of different designs prior to fabrication.

**Acknowledgments**

This study is supported by research funding from the Agency for Science, Technology and Research (A*STAR), Singapore under Grant No. A1820g0084.